# Bridging the Protection Gap: Innovative Approaches to Shield Older Adults from AI-Enhanced Scams

LD Herrera
*Beacom College*
*Dakota State University*
Madison, USA
0009-0005-8027-008X

London Van Sickle
*Beacom College*
*Dakota State University*
Madison, USA
0009-0008-1516-1997

Ashley Podhradsky
*Beacom College*
*Dakota State University*
Madison, USA
0009-0003-9414-707X

*Abstract*—Artificial Intelligence (AI) is rapidly gaining popularity as individuals, groups, and organizations discover and apply its expanding capabilities. Generative AI creates or alters various content types including text, image, audio, and video that are realistic and challenging to identify as AI-generated constructs. However, guardrails preventing malicious use of AI are easily bypassed. Numerous indications suggest that scammers are already using AI to enhance already successful scams, improving scam effectiveness, speed and credibility, while reducing detectability of scams that target older adults, who are known to be slow to adopt new technologies. Through hypothetical cases analysis of two leading scams, the tech support scams and the romance scams, this paper explores the future of AI in scams affecting older adults by identifying current vulnerabilities and recommending updated defensive measures focusing the establishment of a reliable support network offering elevated support to increase confidence and ability to defend against AI-enhanced scams.

## I. Introduction

THE emergence and widespread availability of Artificial Intelligence (AI) offer "extraordinary potential for both promise and peril" [1]. Despite efforts to enact safeguards against malicious use, many modern AI systems, including ChatGPT, have vulnerabilities allowing protective guardrails to be bypassed [2], [3]. Furthermore, some AI tools like FraudGPT or WormGPT were intentionally created without guardrails enabling malicious content generation and expanding the potential for abuse [4].

Scammers equipped with high-quality tools can develop and deploy remarkably effective scams. For example, in 2024, an individual sent $25 million to a scammer because he was fooled by a deepfake video conference [5]. Furthermore, at the 2023 U.S. Senate Special Committee on Aging hearing on AI and Scams, a witness testified that a scammer used voice enhancement to convince the victim to send them money by impersonating their child [6]. These examples suggest scammers are already effectively using AI in scams. According to the Diffusion of Innovation Theory, the adoption of AI in scams is expected to increase once early adopters prove its effectiveness [7].

This work was supported by the NSF NRT Award 1828302.

Older adults are particularly vulnerable to scams due to an increased likelihood of physical or cognitive impairment, which may lead to impaired decision-making abilities [8], [9]. With fewer protective measures, greater accumulated wealth, and a reluctance to report fraud due to the perceived shame, older adults are not only more vulnerable but also more desirable targets [10]–[12]. As the population of adults aged 65 or older grows [13], opportunities for scammers increase, correlating to a rising trajectory of victimization [14].

This paper examines the current and future landscape of AI-enhanced scams targeting older adults by identifying common scam components, exploring potential AI enhancements, and recommending updated defensive strategies.

The following research questions guide this work:

RQ1: What components are frequently found in scams targeting older adults?
RQ2: How will AI enhance these components?
RQ3: What defensive measures are needed against AI-enhanced scams targeting older adults?

The rest of the paper is organized in the following manner: Section 2 presents related works, Section 3 identifies the methodology, Sections 4 provide the study's results divided into five stages, Section 5 discusses the results, implications, limitations, and future research, and Section 6 concludes with an overview of the research.

## II. Related Work

Though no known papers directly explore the future of AI-enhanced scams while focusing on challenges associated with protecting older adults, there is significant research in related topics including Scams Affecting Older Adults, AI Enhancements, and Defensive Updates.

### A. Scams Affecting Older Adults

There is a growing body of research which explores the characteristics of older adults which may make them susceptible to scams. Often, these studies examine the psychological and cognitive factors. They find that cognitive decline, social isolation, and a greater tendency to trust can make older adults particularly susceptible to scams [8], [9]. But other characteristics also contribute to their vulnerabilities including limited awareness of threats, low digital literacy, slow acceptance of new technologies, and limited connections to others who can competently provide support [15]. With increased vulnerabilities, they are often targeted by scammers [16]. Many of these scams are listed in AARP's "Fraud Resource Center" [17].

To respond to this growing concern, the U.S. Federal Trade Commission (FTC) and Department of Justice (DOJ) are tasked with producing annual reports identifying the most costly and frequent scams affecting older adults [18], [19]. While victim reporting remains extremely low, losses for the reporting year 2022-2023 are projected to be up to $48.4 billion [18].

*B. AI Enhancements*

Through advancements in social engineering, automation, and the exploitation of limited human abilities to detect false videos, images, or sounds, AI enhancements increase the frequency, potency, and effectiveness of existing scams [20], [21]. AI enhancements also introduce new threats, including voice and video impersonation [22], [23]. While AI-enhanced or generated artifacts typically have detectable characteristics [24], recent advances reduce the likelihood of detection [25]. Combined with the ability to create "dynamic AI personas that can evolve in real-time" [26] and the ability to deceive [27], AI offers strong incentives to be included in future scams.

*C. Defensive Updates*

Though there is no universal recommendation of what should be done to defend against AI-enhanced scams, Brundage, Avin, Clark, *et al.* [22] and Park, Goldstein, O'Gara, *et al.* [27] recommend prevention and mitigation through responsible development and informed policy-making and legislation. On the other hand, Cross [28] identifies how deepfakes and AI-generated content can bypass the traditional fraud detection method, reverse image searches, potentially giving victims a false sense of security, and advises updates to preventative messaging. Some opt for a more technical approach seeking to counter AI-enhanced threats with AI-enhanced defenses [29], [30], while others suggest normal people should not be expected to become detection experts, rather they should simply default to disbelieving what they see [31].

*D. Gaps Addressed by This Work*

Despite the extensive research on AI enhancements and the rising threat they pose, there is a noticeable gap in defensive strategies for vulnerable populations, such as older adults. Current defenses largely focus on self defense through awareness and technological solutions [32], [33]. This does not consider the shrinking social network [34], resistance to adopt new technologies [35], nor the physical or cognitive conditions that may limit the capacity to recognize scams [36]. This paper aims to fill this gap by developing specific strategies that account for the cognitive, emotional, and technological challenges faced by older adults. The proposed solutions not only focus on enhancing awareness and detection but also emphasize the importance of support systems to provide a comprehensive defense against AI-enhanced scams.

III. METHODOLOGY

This research uses the following five-stage methodology:
- Stage 1: Scam Anatomy
  In response to RQ1, explore common components of scams targeting older adults.
- Stage 2: AI Enhancements
  In response to RQ2, identify how components from the previous stage can be enhanced with AI.
- Stage 3: Hypothetical Cases
  Create hypothetical cases for popular scams using AI-enhanced components identified in Stage 2.
- Stage 4: Case Analysis
  Analyze hypothetical cases defined in Stage 3 exploring gaps in current defenses.
- Stage 5: Recommendations
  In response to RQ3, based on gaps from Stage 4, recommend updated defensive measures.

IV. RESULTS

*A. Stage 1: Scam Anatomy*

Scams are situations in which an adversary, known as a scammer, uses deception to cause another individual, known as a victim, to do something desired by the scammer. Often the goal of a scammer is to convince the victim to provide the scammer with financial benefits or information. This study explored 84 scams known to victimize older adults identified by AARP [17], the FTC [18], and IC3 [19]. Each scam was dissected into parts which were grouped and categorized as shown in Figure 1. The top layer consists of two categories: Protection Factors and Scam Components.

Protection Factors are elements that may limit or prevent scams from achieving their goals. Often these factors include some form of detection, prevention, and/or mitigation. These factors include but are not limited to awareness, experience, warning signs, technical and procedural controls, other people, physical and cognitive abilities, desire to believe, and caution. When protection factors offer inadequate protection, vulnerabilities exist.

Scam Components are elements employed to achieve scam goals. In response to RQ1, the following scam components are frequently found in scams affecting older adults and can be separated into three sub-categories: support materials, communications, and processes. These subcategories are defined in the following manner:
- Support Materials (SMs) serve as the foundation for scams and consist of Background Information, Technological Elements, Physical Materials, and some forms of Communications.
  - Background Information may include descriptions of people (family, friends, boss, authority figure), places

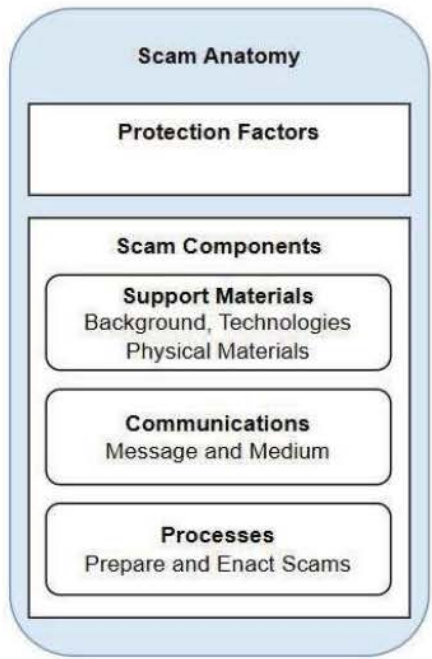

Fig. 1. Visualization of scam anatomy.

(city, country, address), events (work history, educational history, social history, family history), service offerings (help desk, attorney, police officer, etc.), or information about the victim or something/someone the victim cares about.
- Technologies may include phone numbers, emails, IP addresses, social media profiles, websites, mobile apps, or digital ads.
- Physical Materials may include paper products (business card, pamphlet, brochure, mailing label, check, cashier check, invoice, contract, forms), ID badges, tools or equipment, places (office, store, house, apartments, or building), signage including QR codes, costumes, vehicles (cars, vans, motorcycles, boats), humans, animals, or other items that exist in the real world.
• Communications are messages conveyed between scammer(s) and victim. Though there are many communication models, for this discussion, a simplified version is employed where communications consist of only two parts. The first part is the message, including both literal and contextual meaning. The second part is the medium which often taking the form of text, images, audio, video, or in-person discussions.
• Processes are activities that prepare and enact communications and support materials and may include design, training, victim targeting, reconnaissance, initial contact, support material design, development, implementation, or analysis.

### B. Stage 2: AI Enhancements

This stage examines how AI is expected to enhance several of the scam components identified in the previous stage. These expectations are not speculative but are instead projections based on two key trends. The first key trend is that illicit organizations often follow traditional business practices to achieve common business objectives such as increasing profits and minimizing risks [37]. These goals are often achieved through the implementation of new technologies such as AI [38]. In fact, a recent survey showed that 85% of companies ranked AI implementation as was one of their top-five priorities [39]. However, scammers are not necessarily expected to adopt new technologies quickly. The diffusion of innovations theory suggests new technologies are adopted at different rates ranging from early adopters to laggards [40]. Another way organizations achieve their goals is through targeting, which brings a product or service directly to the most suitable individual [38]. These are not the only similarities between traditional organizations and scammers, but they can help guide the understanding of future threats.

The second trend that plays a significant role is the continued advancement of AI technologies. Some of these advancements are a result of the competition between AI service providers seeking to generate undetectable artifacts and others who seek new detection methods [23]. At the same time, human detection abilities remain unchanged leaving many who do not have access to high-end detection systems unable to detect AI enhancements. This scenario is similar to the detection competition between computer viruses and antivirus software that led to a reduction of signature-based detection and an increased focus on behavioral-based detection methods [41]. Furthermore, AI systems can currently be used in an AI-powered scam tactics including the generation of messages, audio, and video, combined with deception strategies [5], [42]. Together, these trends suggest future scams will exploit advanced technologies, increase targeting of vulnerable populations, and reduce detectability for those without access to sophisticated detection systems [43].

Based on these trends, AI enhancements are likely to have the following effects on scam components:

*1) Communications:* AI's generative capabilities will directly impact four types of communications: text, images, voice, and video.

*a) Text:* Text is frequently used to communicate with victims through emails, texts, messaging systems, websites, web apps, social media posts, paper documents, among others. AI enhancements enable text to be altered or replaced, obscuring scammer characteristics such as language fluency or vocabulary selection, while employing customized deceptive methods to achieve goals [27], [44].

*b) Images:* While holding a reasonable amount of scrutiny due to decades of photo editing with software like Photoshop [24], images still hold their place as evidence of truth. Yet, AI enhancements will not only modify existing images but also fabricate new ones where detectable charac-teristics are minimized leading to the proof of people, places, events, and activities that never existed.

  *c) Voices:* Voices will be altered or completely replaced and may have different accents, inflections, sex, and even different languages. AI enhancement may remove characteristics that bio-metrically identify voices or background noises. In some situations, such as live voice alteration, the original words remain, but in other situations, AI will determine what is said.

  *d) Videos:* Videos have traditionally offered a high level of assurance that people are who they purport to be. AI will enable complete replacement or alteration of the visual and audio components of video communications including full body deepfakes [45].

 2) *Background:* At the heart of every scam is a story or pretext that establishes the validity of the scam. Weak stories are difficult to believe and easy to detect. Scammers increase the strength of their stories through experience or by customizing them to fit the situations at hand [46]. AI enhancements provide comprehensive customized stories using a deeper understanding of situations, information gathered from the internet including deep and dark web intelligence, the target's needs, and the best approaches to satisfy those needs while achieving scam goals.

 3) *Technologies:* Technologies exist that offer validation of circumstances and situations, such as caller IDs identifying the origin of calls, emails identifying senders, and websites including those that take payments purport to be true and trustworthy providers and processors of information. Many of these technologies are easily manipulated. Even so, attacks continue to find success as the inherent trust in technology overcomes suspicion [47]. In an AI-enhanced future, scams will continue to exploit technological vulnerabilities, but their likelihood of maneuvering through defenses will increase due to a reduction in flaws that trigger discovery. Furthermore, the pervasiveness of technology will offer new exploitation opportunities.

 4) *Physical Materials:* Physical materials that lend credibility to a scammer's story generally offer a higher level of believability [48]. This is especially true when the victim can interact directly with supporting materials through not only sight and sound but also through other senses. AI enhancements to physical materials include rapid customization and high-quality design of physical materials used as props or supporting evidence.

 5) *Processes:* Scams are enacted through processes including victim targeting, reconnaissance, generating or designing support material, and scammer training. This process has not changed much in thousands of years; however, AI enhancements to these processes will increase the effectiveness and likelihood of success. For instance, AI will be able to assess past results to determine the characteristics of the best victims. It will use those characteristics to find new targets. AI will also be able to guide scammers in selecting the best moves necessary to beat their ill-equipped human opponents. AI will also be capable of executing some scams autonomously, continuously refining its techniques based on past successes and failures using machine learning.

## C. Stage 3: Hypothetical Cases

According to [19], the Tech Support and the Romance scams are two of the leading internet crimes associated with older adults. These scams are used in the hypothetical cases presented below to increase familiarity and meaningfulness to the reader while exploring scam components that are commonly used in other scams affecting older adults, identified in Stage 1, and their AI enhancements, identified in Stage 2.

 1) *Case I: Tech Support Scam:*

  *a) Background:* William is an older retired gentleman who lives alone and uses his home computer to read the news. He has used computers since the 1990s but has limited access to technical support now that he is retired.

  *b) Scam:* While reading news online, a popup message appeared on William's screen indicating that his computer was infected. The message instructed him to call an 800 number for assistance and appeared to be like system messages he had seen in the past. When he tried to close the message, it would reappear with an annoying beep.

  Out of frustration, William calls the number and an AI-generated professional-sounding technical support representative named 'Alice' answers with an accent matching his own. Alice directs William to an official-looking website where, following her instructions, he downloads and runs remote access software. William is directed to a payment page. Alice persuades William to purchase a multi-year plan service plan. Alice downloads and installs additional software removing the popup message and William is happy.

 2) *Case II: Romance Scam:*

  *a) Background:* Janet, a recent widow, is suffering from depression and has turned to support groups on Facebook to communicate with others who are also enduring grief after losing loved ones.

  *b) Scam:* Through Facebook, Janet meets 'Sam', an AI-generated persona of a handsome and wealthy older man who recently lost his wife. He mirrors her emotional state and shares details about himself with supporting material including photos and videos. Sam moves communication away from Facebook to more private channels like WhatsApp, phone calls, and video chats where the relationship turns into a romance.

  Sam claims to be frequently out of the country on business as the manager of a construction company, helping a small community in a foreign land. After he completes his work, he tells her they will be together forever. Suddenly, there is an emergency, and he needs money to complete the final stage of the project. She is suspicious and lets him know her concerns. He provides pictures of his driver's license, passport, and photos of the invoice that he needs to pay to finish the job. The pictures look real to her. She performs reverse image searches on all images and there are no hits. She even has friends look and they agree the pictures look authentic, but they caution her. She ignores their warnings and wires the money. She continues financially support Sam until the scam is terminated.

### D. Stage 4: Case Analysis

This section explores the effectiveness of current defensive measures against the hypothetical AI-enhanced scams and includes a description of relevant victim characteristics, AI enhancements, current defenses, and a summary of defensive gaps.

*1) Analysis of Case I: Tech Support Scam:*

  *a) Victim Characteristics:* Like many older adults, William spends much of his time alone and relies on others for assistance [49]. Older adults tend to be slow to adopt new technologies and, often rely on older technologies [35].

  *b) AI Enhancements:* This case involves two AI-enhanced scam components that significantly affect the quality, detectability, and performance of the scam. The AI-Generated Tech Support Representative, "Alice", is a virtually undetectable fabrication of AI voice generation and textual content generator customized to increase trust and establish familiarity by matching William's language and dialect while effortlessly using persuasion to deceive William and achieve goals [46], [50]. Additionally, the AI-Generated Website matches the look and feel of traditional technical support providers with multiple pages, professional design, and accurate use of language matching William's expectations, reducing apprehension and increasing trustworthiness [51].

  *c) Evaluation of Current Defenses:* While there is likely endpoint protection that can protects against malicious pop-ups, that protection was not in place in this example because home computers typically do not have robust technological defenses [52]. The current FBI and AARP guidance on the tech support scam focuses on situational awareness and detection of characteristics indicative of fraudulent activities while including general suggestions such as "Don't ever call the phone number on a pop-up" and "Don't let an unknown, unverified person get into your computer" [53], [54]. The guidance may have been helpful, but in this case, like many older adults, William was neither aware of the scam nor the general suggestions. But even if he had been provided this information through recent awareness training, it is possible that, due to his advanced age, he may have had troubles remembering that information. William also been taught that scammers tended to be foreign [55]. Alice's custom voice and dialect remove this potential flag and lowers his defenses due to the perceived familiarity [46]. It is also common for individuals experiencing technical difficulties to follow instructions on the screen and to reach out to technical support for assistance. This scam led William to follow the actions he thought to be most appropriate and used the website and Alice to reinforce his beliefs. At the conclusion of the scam, William did not even know he had been scammed. No reporting or investigation followed.

  *d) Defensive Gaps:* Current defenses are unlikely to effectively detect and protect William from this scam. Like many older adults, William lacks connections to others who he can rely on for technical support. Additionally, he does not access to detect potentially fraudulent elements. Finally, he has outdated and ineffective awareness training and due to his inability to recognize the scam, the scam was not reported to authorities.

*2) Analysis of Case II: Romance Scam:*

  *a) Victim Characteristics:* Janet is suffering grief and depression following the loss of her spouse, leaving her emotionally vulnerable to romantic overtures. Once romantically captivated, her desire to continue the romance overcomes concerns of potential problems, reducing skepticism an increasing the likelihood she believes Sam's story [56].

  *b) AI Enhancements:* This case involves three AI-enhanced scam components. The AI-Generated Romantic Interest, "Sam", is a fabrication of AI voice, textual content, and deepfake video generated with an extensive background that establishes trust through demonstrations of care and compassion. [46], [50]. Sam's AI-Generated Facebook profile is constructed matching characteristics of legitimate Facebook profiles having many posts and friend connections to reduce suspicions and increases trustworthiness. Finally, AI-Generated Images form the core of Sam's claims as part of the Facebook profile and as images of documents substantiating Sam's stories.

  *c) Evaluation of Current Defenses:* Through text-based, audio, and visual conversations, Janet was unable to detect his true state which, if discovered, may have provided powerful evidence that she was being scammed. In this case, Janet knew a little about romance scams, perhaps due to a recent influx of romance scam news reports [14]. And because of her initial suspicions, she sought additional guidance, but most of the guidance relied on her decision-making abilities to determine is she was being scammed [57], [58]. She also attempted the recommendation that was not subjective, reverse image searches, but because the images were AI generated and used only for her, no warnings were triggered. She also examined Sam's Facebook profile, but without knowing what to look for, she was easily satisfied. Janet's desire to maintain the relationship was very powerful and weighed heavily in her evaluation using official guidance. Her age may also contribute to cognitive challenges. Still, she took an approach that many others take when they have concerns, she turned to others. In this case, the others she turned to were not qualified to perform proper investigations. Unfortunately, their false negatives reduced suspicion.

  *d) Defensive Gaps:* Current defenses are unlikely to effectively detect AI-Enhanced scam components or protect victims from this scam. Public guidance on the romance scam requires victims to have sound judgement and decision-making, but these capabilities may be compromised by being under the influence of a scammer or by age-related issues. Public guidance also recommends a technological evaluation method that is no longer a significant indicator of scam. Furthermore, while advanced AI detection tools may exist, they are unlikely to be infallible and are often unavailable to older adults who are slow to adopt new technologies. Finally, there is a lack of connections with others who can provide elevated detection capabilities.

*E. Stage 5: Recommendations*

The previous section identified four defensive gaps:

G1: A lack of connections to individuals who can provide an elevated level of support.
G2: Current guidance focuses on equipping potential victims with knowledge to protect themselves.
G3: Inadequate motivation to report cybercrimes leads to an incomplete understanding of threats.
G4: Outdated, ineffective, or inaccessible detection systems.

The following recommendations present a novel approach by specifically targeting the identified gaps, while drawing on familiar areas of intervention. The strategy focuses on three key aspects: increasing support for individuals with diminished support groups (addressing G1 and G2), enhancing motivation to report crimes (addressing G3), and improving the capabilities of support groups (addressing G4). Unlike current approaches that place a significant burden of protection on potential victims, these recommendations aim to redirect that burden. They are divided into four interconnected areas: Awareness Training Programs, Legislation and Policy, Recovery, and Technological Advancements.

*1) Awareness Training Programs:*

*a) Objective:* Increase support available to individuals whose support groups have diminished in size and capability addressing G1 and G2, and increase motivation to report crimes addressing G3.

*b) Considerations:* Awareness training provides a significant reduction in scam susceptibility in business environments and for older adults [59], [60]. The traditional approach empowers older adults to protect themselves through awareness of threats and is most effective when regularly reinforced [61]. Older adults tend to have limited motivation for regular reinforcement [62]. Furthermore, physical or cognitive disabilities may limit their ability to employ awareness knowledge [36].

*c) Strategy:* Increasing training opportunities should provide positive results for many older adults, however, awareness should be adjusted to focus less on protecting oneself and more on reaching out to others who are better equipped to recognize advanced scams. This involves increasing the number of training opportunities through collaboration with adult service providers including healthcare, religious, financial, and non-profit organizations; establishing a 24-hour call center to answer questions and serve as centralized repository for reporting suspicious activities; and using investigations and reports to update training material. This recommendation requires organization between partners at national, state, and local levels.

*2) Legislation and Policy:*

*a) Objective:* Increase support available to individuals whose support groups have diminished in size and capability addressing G1.

*b) Considerations:* Existing laws and regulations provide some protections to older adults, however, technological advancements have altered the threat landscape.

*c) Strategy:* Focuses on the establishment of authority, direction, and funding of a centralized scam defense organization as described in the previous section. Base funding on a percentage of annual projected losses from the previous year so as needs change, funding is adjusted proportionally.

*3) Recovery:*

*a) Objective:* Increase motivation to report crimes addressing G3.

*b) Considerations:* Scams are typically measured by the amount of financial loss, but scams also cause emotional, psychological, physical, social, legal, and reputational harms [63]. But one of the key reasons for not reporting is that there is a lack of confidence that reporting will provide any benefits. For a victim, a road to recovery can essential in providing motivation to report.

*c) Strategy:* This strategy focuses on the use of recovery as a motivator to increase reporting. This can be performed by reinstating tax deductions for scam-related losses that were reported promptly (within 1 week) to law enforcement, modifying Medicare coverage to include support group coverage when facilitated by a Medicare-approved provider and when the crime has been promptly (within 1 week) to law enforcement. Furthermore, it should be shared that quick reporting is the key to financial recovery because bank wires over $50,000 USD can be reversed in a 72-hour window [64].

*4) Technological Enhancements:*

*a) Objective:* Increase the capabilities of support groups addressing G4.

*b) Considerations:* The growing use of AI in everyday communication may diminish the link between AI-related communications and fraudulent activities potentially leading AI detection to be irrelevant, but before that happens, detection is an essential tool that may have stopped the hypothetical cases in their early stages. It should also be remembered that older adults are slow to adopt new technologies [35].

*c) Strategy:* Technological enhancements may not directly assist older adults due to their slow adoption rates, but AI-enhanced protective technologies should continue to be developed while ensuring that new products are easy to use and continuously aware of new scams [65], [66]. Though it is not expected that these new products will be universally accepted, access should be available to those who belong to support groups helping older adults. It is also important to note there are powerful new technologies such as Android's AI-powered listening system that may be significantly more effective than traditional detection technologies [67]. Given the potential for AI detection to be irrelevant, this approach should be given substantial consideration. Also note that this strategy does not attempt to address the slow adoption rate beyond making products easy to use. Protecting older adults should not require constant and continuous technological investment.

## V. DISCUSSION

This paper set out to answer three research questions. A five-stage methodology was employed to respond to these questions. RQ1 sought to uncover commonalities in scams affecting older adults. Through a detailed examination of 84 AARP identified scams, a Scam Anatomy was created in Stage 1 including a common methodology and descriptions of included components answering RQ1. This breakdown of components widely considers numerous scams rather than focusing on a specific scam type.

With these components identified, RQ2 questioned how AI could be used to enhance scams. In Stage 2, the research explored the existing capabilities of AI as well as those capabilities that have not yet come to fruition. This was performed by extrapolating the evolutionary trends of AI development rather than fictional speculation. These capabilities were applied as enhancements to the previously identified components to answer RQ2.

To achieve the final goal of identifying altered protection methods accounting for future AI enhancements, RQ3 required the ability to peer into the future identifying gaps when current defenses are used against future attacks. To accomplish this, hypothetical cases were created in Stage 3, analyzed in Stage 4 to identify gaps. Then, in Stage 5, recommendations responsive to incoming threats were established to increase protections against future threats while mitigating damages experienced by older adult victims.

Background research into the recommended defenses against AI enhancements is filled with novel technologies to increase detection, but these new technologies only work for those who utilize them. As previously mentioned, older adults are likely to continue to be slow in adopting new technologies. Additionally, organizations such as AARP and government organizations such as the FBI and FTC seem place a significant amount of the burden of protection on the potential victims by offering awareness training and recommending best practices to minimize risk. This paper does not discount the value of these protection methods, however, it takes a different direction which is more in line with Kropczynski, Aljallad, Elrod, *et al.* [68], focusing on the establishment of a reliable support network offering elevated support. The increased availability of elevated support will increase confidence and ability to defend against AI-enhanced scams Morrison, Coventry, and Briggs [15].

### A. Implications

This research highlights several important implications for protecting older adults against future AI-enhanced scams. First, it brings attention to the significant damages caused to older adults by scammers, both financial and non-financial. Second, it draws attention to the changing threat landscape and the potential for AI to increase the effectiveness of scams. Third, it identifies gaps where the current defenses may fail against future AI-enhanced scams, highlighting the need to alter defenses. Finally, it offers recommendations to address the gaps specifically focusing on increasing the available support systems available to older adults through a multifaceted approach involving technological innovation, policy reform, education, and community support. These adjustments could generate significant economic and societal impact by reducing the effectiveness of scams.

### B. Limitations and Future Research

This research relies on projections of how AI enhancements could enhance scams targeting older adults. Though these projections are based on the ongoing evolution of AI, anecdotal evidence of AI enhancements already in use, and the expectation that scammers will continue to use technology to overcome victim defenses, neither the future capabilities of AI nor the effectiveness of proposed recommendations can be measured or evaluated against future threats. Only time will provide the opportunities for validation and measurement.

Future research may include the evaluation of the presented recommendations on a small scale as they are applied to existing threats. An alternative approach could be to broaden this research beyond the focus of protecting older adults, extending it to a larger population. As AI enhancements evolve, individuals in all populations may become increasingly susceptible to AI-enhanced scams.

## VI. CONCLUSION

Through the analysis of 84 scams and two hypothetical cases, this paper explores how AI may be used in future scams and identifies gaps in current defensive strategies. It then recommends updates to four areas of defense, including a notable change in direction.

Currently, the identification of fraudulent support materials is key to the identification of a scam. However, the capabilities of AI have reached a point where humans have difficulty distinguishing between real and fake. These capabilities may continue advancing until even AI-powered detection systems struggle. AI enhancements will also become commonplace. A recent study shows 70% of respondents already use filters on their social media posts [69]. Together, these developments suggest that protecting older adults should no longer focus on the detection of false artifacts but rather on the detection of scams at a higher level by increasing situational awareness and understanding of how scams are executed.

Increasing awareness among older adults is challenging. Many face limited motivation, high resistance to change, restricted access to awareness resources, or limited technical skills. While efforts to increase awareness may succeed to some extent, characteristics associated with age including physical, cognitive, or psychological issues may limit the effectiveness of awareness training programs. Additionally, an AI-powered awareness system that listens to or observes situations as they occur, such as the previously mentioned Android system, may provide an essential layer of protection for vulnerable older adults. However, due to the slow technology adoption rate of older adults and the need to overcome privacy concerns, such a system may not be widely accepted. Both technology and awareness are strong solutions that can be very effective for younger populations. For older adults, there needs to be a focus on increasing access to individuals who are both knowledgeable in scam characteristics and able to work with victims. This additional layer of protection should lead to better outcomes for vulnerable older adults using a system that harnesses the collective knowledge of scams occurring across the country or even across the globe.